\documentclass[10pt,twocolumn,letterpaper]{article}

\IfFileExists{spconf.sty}{\usepackage{spconf}\usepackage{amsmath}}%
                         {
\usepackage[letterpaper,top=1.0in,left=0.75in,textwidth=7.0in,textheight=9.0in,columnsep=0.3in]{geometry}

\makeatletter
\newcommand{\name}[1]{\gdef\@name{#1}}
\newcommand{\address}[1]{\gdef\@address{#1}}
\name{}\address{}

\renewcommand{\maketitle}{%
  \twocolumn[%
    \begin{center}
      {\Large\bfseries\@title\par}\vspace{1.2em}
      {\normalsize\@name\par}\vspace{0.4em}
      {\small\itshape\@address\par}\vspace{1.4em}
    \end{center}%
  ]%
}
\makeatother

\renewenvironment{abstract}{\noindent\textbf{Abstract---}\itshape}{\par\vspace{0.6em}}
\providecommand{\keywords}[1]{\noindent\textbf{Index Terms---}#1\par\vspace{0.8em}}

\setlength{\parindent}{1em}
\setlength{\parskip}{0pt}
}

\usepackage{amsmath,amssymb,bm}
\usepackage{graphicx}
\usepackage{booktabs}
\usepackage{multirow}
\usepackage[table,dvipsnames]{xcolor}
\usepackage{url}
\usepackage{cite}
\usepackage{hyperref}
\usepackage{etoolbox}
\patchcmd{\thebibliography}
  {\settowidth}
  {\setlength{\itemsep}{0pt}\setlength{\parskip}{0pt}\settowidth}
  {}{}
  
\newif\ifdraft\drafttrue
\definecolor{DummyRed}{rgb}{0.80,0.10,0.10}
\ifdraft
  
  \newcommand{\todo}[1]{\textcolor{DummyRed}{[\textbf{TODO:} #1]}}
\else
  
  \newcommand{\todo}[1]{}
\fi

\newcommand{\Lc}{\mathcal{L}}
\newcommand{\sat}{s}
 
\title{CORD-KWS: Calibrated, Order-Aware Detection for Open-Vocabulary \\ Keyword Spotting}

\name{Ramesh Gundluru$^{1}$, Adarsh Arigala$^{2}$, Sri Rama Murty Kodukula$^{1}$ }

\address{
$^{1}$ Speech Information Processing Lab, Indian Institute of Technology Hyderabad, India \\
$^{2}$ SPRING Lab, Indian Institute of Technology Madras, India \\
ee22m24p000001@iith.ac.in, arigalaadarsh780@gmail.com,  ksrm@ee.iith.ac.in
}

\begin{document}
\maketitle

\begin{abstract}
Open-vocabulary keyword spotting (KWS) must detect arbitrary keywords without retraining. Cross-attention-based models achieve state-of-the-art performance but require pairwise interaction between the audio and each keyword, recomputing the fused representation for every audio--keyword pair. Embedding-based models avoid this computational cost through independent encoding and similarity scoring, yet remain inferior on challenging benchmarks such as LibriPhrase-hard. We show that this gap is primarily a training issue rather than a limitation of model capacity. While contrastive learning encourages correct keywords to rank above negatives, it does not explicitly calibrate absolute similarity scores, which are critical for applying a fixed threshold across keyword detection. We propose Calibrated, Order-aware Detection KWS (CORD-KWS), a training framework that introduces two complementary objectives while retaining the same encoders, cosine scoring, and $O(1)$ keyword enrollment: a calibrated detection head that supervises absolute scores, and a frame-level CTC objective that preserves phonetic order information. CORD-KWS achieves EERs of $0.43\%$ and $9.64\%$ on LibriPhrase-easy and LibriPhrase-hard, respectively, outperforming existing embedding- and cross-attention-based models.


\end{abstract}

\keywords{Keyword Spotting, Contrastive Learning, Calibration, CORD-KWS}

\section{Introduction}
\label{sec:scope}
Existing open-vocabulary keyword spotting (OV-KWS) systems fall into three families. \textit{Posterior-based} methods use an Automatic Speech Recognition (ASR) model to produce
frame-level posteriors and decode keywords by sequence search
\cite{chen2014small,wang2017small,jin24d_interspeech,kim25d_interspeech}.
\textit{Cross-attention-based} methods model the audio--keyword interaction directly, through cross-attention and a detection classifier \cite{baseline_cmcd,baseline_phonmatchnet,baseline_emkws,ai24_interspeech,plcl,SLiCK}. This fine-grained interaction enables high accuracy, but the audio and keyword representations are jointly computed. As a result, the model must perform this interaction again for each keyword, making the computation grow linearly with the number of keywords, $O(N)$. In contrast, \textit{Embedding-based} methods encode audio and text separately into a shared space and perform detection using cosine similarity
\cite{baseline_triplet,baseline_clad,baseline_adml,synapspot,mate}. Since the acoustic representation is computed only once and can be reused for different keyword queries, embedding-based systems are computationally efficient and particularly suitable for large-scale retrieval and streaming applications. On LibriPhrase-hard dataset,  the best published embedding-based system reports $20$--$30$\%
EER~\cite{baseline_adml}, against $9$--$12$\% for the strongest cross attention based systems~\cite{ai24_interspeech,plcl,SLiCK}. We report both families in Table~\ref{tab:main},
while keeping them in their respective computational cost classes. This comparison highlights the performance gap between the two approaches without implying that an $O(1)$ embedding-based system and an $O(N)$ cross-attention matcher provide equivalent accuracy or computational cost.

\begin{figure*}[tb]
\centering
\includegraphics[width=1\linewidth]{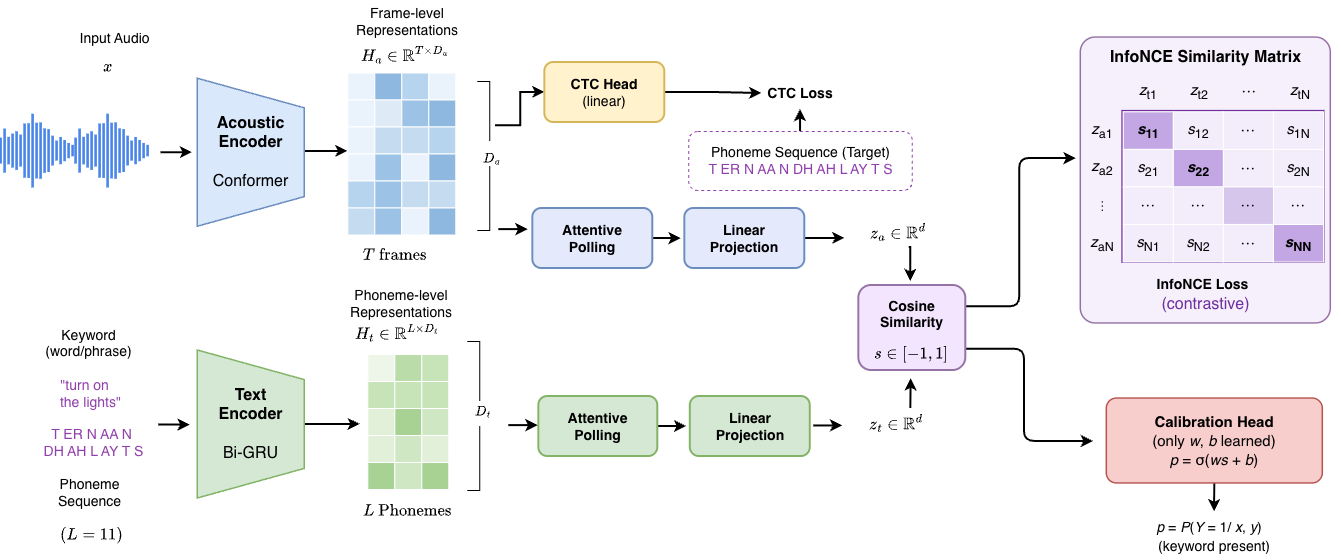}
\caption{Architecture diagram of the proposed CORD-KWS.}
\label{fig:overview}
\end{figure*}

Three limitations of existing embedding-based approaches motivate our design. CLAD~\cite{baseline_clad}, the strongest embedding-based system and the work most closely related to ours, relies on a two-stage and alignment-dependent training procedure.

\textbf{Training is two-stage and alignment-dependent}. CLAD  pre-trains a mono phone acoustic model using forced alignments and then freezes acoustic encoder. During contrastive training, it requires forced alignments again to construct audio negatives based on their overlap with the ground-truth keyword location. Since the acoustic encoder is frozen, the contrastive objective can only adapt the projection heads.

\textbf{Sampled negatives are temporally hard, not phonetically}. CLAD constructs negatives by sampling keywords from other utterances within the batch and by using overlapping windows from the same utterance. Whereas LibriPhrase-hard dataset measures confusability between different words that sound similar.

\textbf{Ranking is not detection.} InfoNCE encourages the correct keyword to have higher \textit{similarity} than incorrect keywords for a given audio segment, but it does not constrain the absolute similarity values. Detection needs exactly one threshold applied to every pair.


We address these issues with \textbf{CORD-KWS}, a calibrated, order-aware
detector. Our contributions are:
\begin{enumerate}
  \item A diagnosis of the ranking--detection mismatch in the contrastive objective, with evidence that absolute score supervision repairs it (Sec.~\ref{sec:method}).
  \item A monotonic calibrated head supplying the supervision; being monotonic, it cannot alter EER by rescaling (Sec.~\ref{sec:head}).
  \item A frame-level CTC auxiliary that imposes phone order, trained jointly and requiring no phoneme level forced alignments and two stage training (Sec.~\ref{sec:ctc}).
  \item Phonetically hard negatives mined by phone edit distance,
        against the \emph{temporally} hard negatives of prior
        work~\cite{baseline_clad} (Sec.~\ref{sec:hardneg}).
  \item State of the art among all the KWS systems: $0.43\%$ / $9.64\%$
        Equal Error Rate (EER) on LibriPhrase easy / hard conditions.
\end{enumerate}
\section{CORD-KWS Methodology}
\label{sec:method}
Fig.~\ref{fig:overview} illustrates the overall architecture of the proposed model. The acoustic encoder takes an audio waveform (x) as input and generates frame-level representations using a Conformer~\cite{conformer} encoder. An attentive pooling layer followed by a linear projection maps the resulting representations into a common latent space, yielding the acoustic embedding $z_a \in \mathbb{R}^{D}$. Similarly, the text encoder processes the phoneme sequence using a BiGRU to obtain phoneme-level representations. An attentive pooling layer and a separate linear projection are then applied to map the pooled representation into the same shared latent space, producing the text embedding $z_t \in \mathbb{R}^{D}$.

Given a batch of $N$ paired audio and text keyword embeddings, we compute the pairwise cosine similarity between each acoustic embedding $z_a^i$ and text embedding $z_t^j$ as
\begin{equation}
s_{ij} =
\frac{z_a^i \cdot (z_t^j)^T}
{\|z_a^i\|_2 \|z_t^j\|_2}.
\end{equation}
The resulting $N \times N$ similarity matrix contains the matched audio-text pairs along its diagonal. We employ a bidirectional InfoNCE objective with temperature $\tau$. The audio-to-text loss is defined as
\begin{equation}
\mathcal{L}_{a\rightarrow t}
=
-\frac{1}{N}
\sum_{i=1}^{N}
\log
\frac{
\exp(s_{ii}/\tau)
}{
\sum_{j=1}^{N}
\exp(s_{ij}/\tau)
}.
\end{equation}
Similarly, the text-to-audio loss is given by
\begin{equation}
\mathcal{L}_{t\rightarrow a}
=
-\frac{1}{N}
\sum_{j=1}^{N}
\log
\frac{
\exp(s_{ii}/\tau)
}{
\sum_{i=1}^{N}
\exp(s_{ij}/\tau)
}.
\end{equation}
The overall contrastive loss is the average of the two losses:
\begin{equation}
\mathcal{L}_{\mathrm{InfoNCE}}
=
\frac{1}{2}
\left(
\mathcal{L}_{a\rightarrow t}
+
\mathcal{L}_{t\rightarrow a}
\right).
\end{equation}

The three additions below supply the missing supervision as mentioned in the Sec ~\ref{sec:scope}.

\subsection{Phonetically hard negatives}
\label{sec:hardneg}

For every training keyword, we mine its $k$ nearest neighbours in the training vocabulary under phone edit distance and sample $n_{\text{hard}}$ per anchor as additional text negatives, appended as extra columns of the InfoNCE denominator in equation (2).

\subsection{Order supervision via a CTC auxiliary}
\label{sec:ctc}

Attentive pooling is permutation-invariant and, therefore, does not explicitly preserve the temporal order of acoustic frames. Moreover, $\mathcal{L}_{\text{InfoNCE}}$ is applied only to the final pooled embedding and provides no direct supervision for phoneme order. So, we add a linear head to the Conformer output before attentive pooling and are jointly trained using $\Lc_{\text{CTC}}$ loss using phoneme sequence of the audio keyword. The CTC head is discarded after training so, the scoring, enrollment, and inference cost are unchanged.



\subsection{Absolute supervision via a calibrated detection head}
\label{sec:head}

Two properties of this objective motivate the formulation. Writing the loss 2 example as $\Lc_i =
-\log \frac{\exp(\gamma \sat_{ii})}{\sum_k \exp(\gamma \sat_{ik})}$, the
substitution $\sat_{ik} \mapsto \sat_{ik} + c_i$ leaves $\Lc_i$ invariant for every $c_i$. The geometry is fixed only up to a per-utterance shift, and a single detection threshold cannot absorb one. To fix, this we map the similarity to a posterior through a affine transform,
\begin{equation}
p = \sigma\!\left(w\,\sat + b\right)
\label{eq:head}
\end{equation}
with $w$ reparameterised as a softplus and we use $\Lc_{\text{BCE}}$ to provide absolute supervision against the
pair label ($y=1$ for matched pairs, $y=0$ for the negatives of
Sec.~\ref{sec:hardneg}).
The total loss is defined as 
\begin{equation}
\Lc_{\text{Total}} = \Lc_{\text{InfoNCE}} + \lambda_{\text{ctc}}\Lc_{\text{CTC}}
      + \lambda_{\text{bce}}\Lc_{\text{BCE}}
\end{equation}
where  $\lambda_{\text{ctc}},\lambda_\text{bce}$ are relative weights for each losses. We swept $\lambda_{\text{bce}}$ and $\lambda_{\text{ctc}}$ one at a time (Table~\ref{tab:weights}) and selected
$(\lambda_{\text{bce}},\lambda_{\text{ctc}})\!=\!(1.0,0.3)$ configuration.

The gradient passes through the
encoders, driving confusable negatives towards $0$ and positive keyword pairs towards $1$ in the shared space. Because $w>0$ makes $\sat \mapsto p$ strictly
increasing, the head preserves every pairwise ordering and cannot change EER or AUROC at inference. 

\section{Experimental Setup}

We use LibriPhrase~\cite{baseline_cmcd} datset, derived from LibriSpeech~\cite{librispeech}. Training
uses the train-clean-100 and train-clean-360 trials; evaluation uses the train-other-500 trials.
We set $d_{\text{model}}=D=64$. The audio encoder consists of four Conformer layers with four attention heads, an FFN dimension of 128, and a depthwise convolution kernel size of 7. The text encoder is a two-layer BiGRU with 64-dimensional input embeddings and 128 hidden units. The deployed model contains 848,k parameters, compared with 2.2,M reported for CLAD; an additional 3k parameter CTC head is used only during training and discarded at inference. We train for 60 epochs with
AdamW~\cite{loshchilov2019adamw} at a peak learning rate of $10^{-3}$ and a batch of 128 per GPU; since in-batch negatives are not gathered across devices, that batch size \emph{is} the negative count. The mining and loss weights are $n_{\text{hard}}\!=\!4$ drawn from the top 32 phonetic neighbours, $\lambda_{\text{ctc}}\!=\!0.3$ and  $\lambda_{\text{bce}}\!=\!1.0$, held fixed across the experiments. We report Equal Error Rate (EER, $\downarrow$), the threshold at which false
accepts equal false rejects, and area under the ROC curve (AUROC, $\uparrow$), both computed over all audio--keyword pairs and reported separately for the easy and hard LibriPhrase conditions.

\section{Results and Discussion}
\label{sec:results}

\subsection{Performance evaluation on LibriPhrase}
\label{sec:capacity}

\begin{table}[t]
\centering \scriptsize
\caption{LibriPhrase results, grouped by interaction mechanism. PT / FT are
pre-training and fine-tuning hours; $*$ ASR-based pre-training, $\dagger$
phrase-based, $\ddagger$ synthetic data. The grouping decides whether the text
side can be cached: \emph{embedding} systems reduce a new keyword to one vector,
$O(1)$ per enrolment, and score by inner product; \emph{cross-attention}
systems fuse the pair and must recompute for every keyword; \emph{posterior}
systems share the acoustic pass but run a per-keyword decoding search. IS : INTERSPEECH and ICP: ICASSP.}
\label{tab:main}
\setlength{\tabcolsep}{1pt}
\begin{tabular}{@{}llcccc c cc c cc@{}}
\toprule
\multirow{2}{*}{\textbf{Method}} & \multirow{2}{*}{\textbf{Venue}} &
\multirow{2}{*}{\textbf{\#P(M)}} &
\multirow{2}{*}{\textbf{PT (h)}} & \multirow{2}{*}{\textbf{FT (h)}} & & &
\multicolumn{2}{c}{\textbf{AUROC (\%)} $\uparrow$} & &
\multicolumn{2}{c}{\textbf{EER (\%)} $\downarrow$} \\
\cmidrule(lr){8-9}\cmidrule(lr){11-12}
 & & & & & & & \textbf{LP}$_\textbf{H}$ & \textbf{LP}$_\textbf{E}$ & &
             \textbf{LP}$_\textbf{H}$ & \textbf{LP}$_\textbf{E}$ \\
\midrule
\multicolumn{12}{@{}l}{\textit{Posterior-based}} \\
DONUT~\cite{lugosch2018donut}              & ICP 21 & ---   & 0        & 460 & & & 62.55 & 78.74 & & 41.95 & 28.74 \\
CTCAT~\cite{jin24d_interspeech}            & IS 24     & 0.2M  & 0        & 460 & & & 77.10 & 98.32 & & 29.63 & 6.06 \\
W-CTC~\cite{kim25d_interspeech}            & IS 25     & 3.6M  & 460$\dagger$ & 460 & & & 95.93 & 99.95 & & 10.21 & 0.91 \\
\midrule
\multicolumn{12}{@{}l}{\textit{Cross-attention-based}} \\
CMCD~\cite{baseline_cmcd}                  & IS 22     & 0.7M  & 0        & 460 & & & 73.58 & 96.70 & & 32.90 & 8.42 \\
EMKWS~\cite{baseline_emkws}                & IS 23     & 3.7M  & 460$\dagger$ & 460 & & & 84.21 & 97.83 & & 23.36 & 7.36 \\
PhonMatchNet~\cite{baseline_phonmatchnet}  & IS 23     & 0.7M  & 110{,}000 & 460 & & & 88.52 & 99.29 & & 18.82 & 2.80 \\
CED~\cite{baseline_ced}                    & ICP 24 & 3.6M  & 460$\dagger$ & 460 & & & 92.70 & 99.84 & & 14.40 & 1.70 \\
MM-KWS@T~\cite{ai24_interspeech}           & IS 24     & 3.9M  & 0        & 887$\ddagger$ & & & 95.36 & 99.94 & & 10.41 & 0.82 \\
SLiCK~\cite{SLiCK}                         & ICP 25 & 0.6M  & 460$\dagger$ & 460 & & & 94.90 & 99.82 & & 11.10 & 1.78 \\
PLCL@T~\cite{plcl}                         & ICP 25 & 40M   & 680{,}000$\dagger$ & 460 & & & 95.56 & 99.95 & & 9.96 & 1.21 \\

\midrule
\multicolumn{12}{@{}l}{\textit{Embedding-based}} \\
Triplet~\cite{baseline_triplet}   & IS 19     & ---   & 0       & 460 & & & 54.88 & 63.53 & & 44.36 & 32.75  \\
CLAD~\cite{baseline_clad}         & ICP 24 & 3.6M  & 460$*$  & 460   & & & 76.15 & 97.03 & & 30.30 & 8.65 \\
ADML~\cite{baseline_adml}         & IS 25     & 1.8M  & 0 & 4600$*$   & & & 88.71 & 99.86 & & 20.09 & 1.33 \\
SynapSpot-TA~\cite{synapspot}     & ICP 26 & 0.9M    & 460       & 460   & & & 79.15 & 97.34 & & 27.29 & 5.77 \\
MATE~\cite{mate}                  & ICP 26 & ---     & 0       & 4600$*$   & & & 88.70 & 99.86 & & 20.06 & 1.38 \\

\textbf{CORD-KWS-tiny }     & ---       & 0.85M & 0 & 460 & & & 95.05 & 99.97 & & 11.46 & 0.57 \\
\textbf{CORD-KWS-small }    & ---       & 2.99M & 0 & 460 & & & \textbf{96.26} & \textbf{99.98} & & \textbf{9.64} & \textbf{0.43} \\

\bottomrule
\end{tabular}
\end{table}

 We evaluate our CORD-KWS at two model scales, with $0.85$M and $2.99$M parameters (Table~\ref{tab:main}). Increasing the model size improves performance on hard negatives, reducing the EER from $11.46\%$ to $9.64\%$, while on easy negatives, improved from $0.57\%$ to $0.43\%$. The small variant model indeed gives lower EER, but our primary goal is to study the effect of the proposed training objectives under a fixed compact architecture. We therefore use the tiny configuration consistently for the ablations.

CORD-KWS substantially outperforms existing embedding-based systems on both hard and easy negatives. ADML~\cite{baseline_adml} and MATE~\cite{mate} report EERs of $20.09\%$ and $20.06\%$ on hard negatives, respectively, whereas CORD-KWS-tiny alone achieves $11.46\%$ on hard and $0.57\%$ on easy negatives. It also outperforms cross-attention and posterior-based systems on both splits. CORD-KWS-tiny uses $4.6\times$ fewer parameters than comparable systems such as MM-KWS@T and W-CTC. CORD-KWS-small achieves state-of-the-art performance by outperforming PLCL@T while using $13.4\times$ fewer parameters and requiring no pre-training. Overall, CORD-KWS-small achieves the best performance in the table on both hard and easy negatives, while retaining the $O(1)$ keyword enrolment and scoring advantage of embedding-based KWS.

\subsection{What drives the gain?}
\vspace{-0.5cm}
\begin{table}[!h]
\centering
\caption{Cumulative ablation on the tiny model ($0.85$\,M). Each row adds one
component to the row above; architecture, batch size, optimiser and data splits
are identical throughout.}
\label{tab:ladder}
\setlength{\tabcolsep}{6pt}
\begin{tabular}{@{}llcc@{}}
\toprule
 & \multirow{2}{*}{\textbf{System}} &
 \multicolumn{2}{c}{\textbf{EER (\%) $\downarrow$}} \\
\cmidrule(lr){3-4}
 & & \textbf{LP}$_\textbf{E}$ & \textbf{LP}$_\textbf{H}$ \\
\midrule
S1 & in-batch neg.    & 0.63 & 14.94 \\
S2 & + phonetic hard  & 0.59 & 13.66 \\
S3 & + CTC auxiliary  & 0.51 & 12.74 \\
S4 & + detection head & 0.57 & 11.46 \\
\bottomrule
\end{tabular}
\end{table}

To understand where the performance gains come from, we perform a cumulative ablation study on the CORD-KWS-tiny model (Table~\ref{tab:ladder}). Starting from in-batch negatives (S1), we progressively add phonetic hard negatives, the auxiliary CTC objective, and the detection head, while keeping the architecture, batch size, optimizer, and data splits fixed. The results show that each component improves hard-negative detection. Phonetic hard negatives reduce EER from $14.94\%$ to $13.66\%$, while the CTC auxiliary loss further reduces it to $12.74\%$. The detection head provides the final improvement to $11.46\%$. The detection-head BCE uses all hard negatives along with only a few easy examples, giving greater emphasis to hard negatives. Consequently, it improves $LP_H$ substantially, with a small increase in $LP_E$ from $0.51\%$ to $0.57\%$.



\begin{table}[t]
\centering
\caption{Loss-weight ablations on the tiny model. Each weight is swept independently around S4, the shared centre of both sweeps. $\lambda_{\text{bce}}\!=\!0$ corresponds to S3 (Table~\ref{tab:ladder}).}
\label{tab:weights}
\begin{tabular}{@{}lcc@{}}
\toprule
 & \multicolumn{2}{c}{\textbf{EER (\%) $\downarrow$}} \\
\cmidrule(lr){2-3}
 & \textbf{LP}$_\textbf{E}$ (Easy) & \textbf{LP}$_\textbf{H}$ (Hard) \\
\midrule
\multicolumn{3}{@{}l}{\emph{Head weight} $\lambda_{\text{bce}}$ (at $\lambda_{\text{ctc}}\!=\!0.3$)}\\
0.1      & 0.53 & 12.21 \\
0.5      & 0.53 & 11.65 \\
1.0 (S4) & 0.57 & \textbf{11.46} \\
\midrule
\multicolumn{3}{@{}l}{\emph{CTC weight} $\lambda_{\text{ctc}}$ (at $\lambda_{\text{bce}}\!=\!1.0$)}\\
0.1      & 0.60 & 11.81 \\
0.3 (S4) & 0.57 & \textbf{11.46} \\
1.0      & 0.57 & 11.70 \\
\bottomrule
\end{tabular}
\end{table}

\subsection{Keyword Localisation in Continuous Speech}

\begin{figure}[]
\centering
\includegraphics[width=\columnwidth]{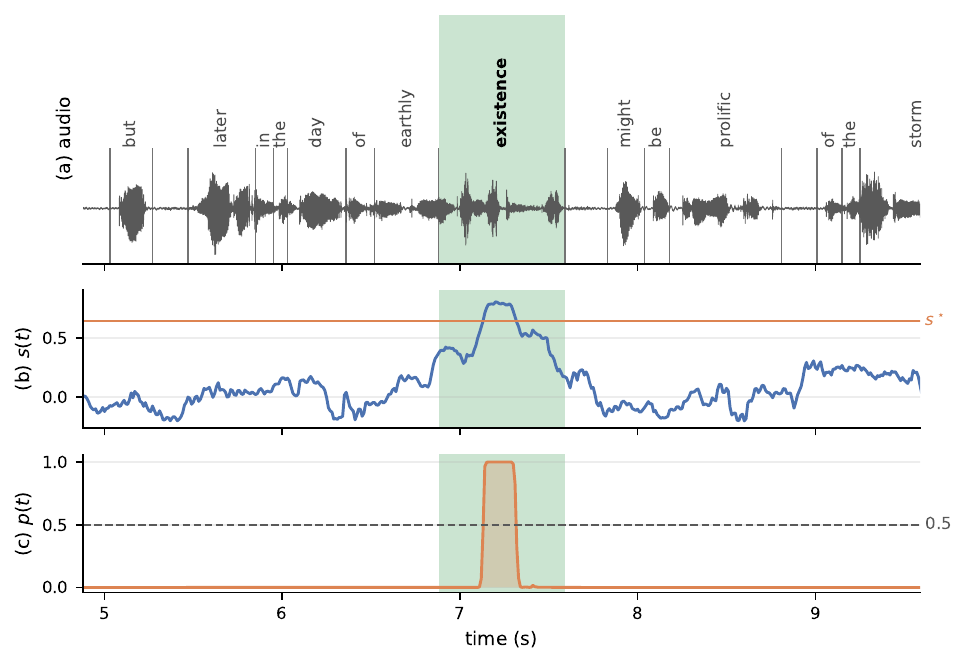}
\caption{Keyword localisation in continuous speech for the query \emph{existence} on LibriSpeech test-clean}.
\label{fig:traj}
\end{figure}

To localize keyword in continuous speech, we slide a 1s window over the utterance with a hop size 10ms. For each window, we extract an acoustic embedding and compute its similarity with the text keyword embedding. Repeating this
process across the utterance produces the temporal similarity trajectory $s(t)$ shown in Fig.~\ref{fig:traj}(b). The similarity scores are subsequently mapped to detection probabilities using the detection head, resulting in posteriors over time Fig.~\ref{fig:traj}(c). The resulting score posteriors show a clear peak around the ground-truth keyword span, while remaining low elsewhere in the utterance. This indicates that the model can localize the queried keyword in continuous speech, even though it was trained only on isolated phrases.

\section{Conclusion}
\label{sec:conclusion}

We show that the accuracy gap between embedding-based and cross-attention-based KWS can be substantially reduced through improved training rather than increased model capacity. CORD-KWS combines phonetically mined hard negatives, an auxiliary CTC objective, and a monotone calibrated detection head while preserving the efficient embedding-based inference with $O(1)$ keyword enrolment. CORD-KWS-small achieves state-of-the-art EERs of $9.64\%$ and $0.43\%$ on the hard- and easy-negative trials, using only $2.99$M parameters and no pre-training. Furthermore, despite being trained only on isolated keywords/phrases, CORD-KWS can localize keywords in continuous speech, demonstrating its potential for efficient keyword localization.

\section{Generative AI Use Disclosure}
We used ChatGPT for linguistic refinement only. All experiments, analyses, and scientific conclusions were conducted and developed by the authors.

\bibliographystyle{IEEEbib}
\bibliography{refs}

\end{document}